\documentclass{article}%
\usepackage{amsfonts}
\usepackage{amsmath}
\usepackage{amssymb}
\usepackage{graphicx}%
\newtheorem{theorem}{Theorem}

\newtheorem{condition}[theorem]{Condition}

\newtheorem{definition}[theorem]{Definition}
\newtheorem{example}[theorem]{Example}

\newtheorem{lemma}[theorem]{Lemma}

\newtheorem{proposition}[theorem]{Proposition}
\newtheorem{remark}[theorem]{Remark}

\newenvironment{proof}[1][Proof]{\noindent\textbf{#1.} }{\ \rule{0.5em}{0.5em}}
\begin{document}

\title{On a relation of pseudoanalytic function theory to the two-dimensional
stationary Schr\"{o}dinger equation and Taylor series in formal powers for its
solutions }
\author{Vladislav V. Kravchenko\\Secci\'{o}n de Posgrado e Investigaci\'{o}n\\Escuela Superior de Ingenier\'{\i}a Mec\'{a}nica y El\'{e}ctrica\\Instituto Polit\'{e}cnico Nacional\\C.P.07738 M\'{e}xico D.F., \\MEXICO\\phone: 525557296000, ext. 54782\\e-mail: vkravchenko@ipn.mx}
\maketitle

\begin{abstract}
We consider the real stationary two-dimensional Schr\"{o}dinger equation. With
the aid of any its particular solution we construct a Vekua equation
possessing the following special property. The real parts of its solutions are
solutions of the original Schr\"{o}dinger equation and the imaginary parts are
solutions of an associated Schr\"{o}dinger equation with a potential having
the form of a potential obtained after the Darboux transformation. Using L.
Bers theory of Taylor series for pseudoanalytic functions we obtain a locally
complete system of solutions of the original Schr\"{o}dinger equation which
can be constructed explicitly for an ample class of Schr\"{o}dinger equations.
For example it is possible, when the potential is a function of one cartesian,
spherical, parabolic or elliptic variable. We give some examples of
application of the proposed procedure for obtaining a locally complete system
of solutions of the Schr\"{o}dinger equation. The procedure is algorithmically
simple and can be implemented with the aid of a computer system of symbolic or
numerical calculation.

\textbf{Keywords}: Schr\"{o}dinger equation, factorization, pseudoanalytic
functions, generalized analytic functions, p-analytic functions, Darboux
transformation, exact solutions

PACS numbers: 02.30.-f, 02.30.Tb, 30G20, 35J10

\end{abstract}

\section{Introduction}

The appearance of a mathematical theory depends a lot on the preferences and
mathematical tastes of its creator. Different people regard and describe the
same mathematical results and ideas from different viewpoints, and this is one
of the sources of the richness and progress of our science.

The pseudoanalytic function theory is one of the clearest confirmations of
this assertion. It was independently developed by two prominent mathematicians
I. N. Vekua and L. Bers with coauthors, and presented in their books
\cite{Berskniga} and \cite{Vekua}. The theory received further development in
hundreds of posterior works (see, e.g., the reviews \cite{Begehr} and
\cite{Tutschke}), and historically it became one of the important impulses for
developing the general theory of elliptic systems. Here the Vekua theory
played a more important role due to its tendency to a more general,
operational approach. L. Bers tried to follow more closely the ideas of
classical complex analysis and paid more attention to the efficient
construction of solutions. Among other results L. Bers obtained analogues of
the Taylor series for psedoanalytic functions and some recursion formulas for
constructing generalizations of the base system $1$, $z$, $z^{2}$,$\ldots$ .
The formulas require knowledge of the Bers generating pair (two special
solutions) of the corresponding Vekua equation describing pseudoanalytic
functions as well as generating pairs for an infinite sequence of Vekua
equations related to the original one. The necessity to count with an infinite
number of exact solutions of different Vekua equations resulted to be an
important obstacle for efficient construction of Taylor series for
pseudoanalytic functions.

Nevertheless as we show in the present work the Bers recursion formulas seem
to have been specially designed for obtaining Taylor-type series for solutions
of two-dimensional stationary Schr\"{o}dinger equations admitting particular
solutions which enjoy a peculiar property called in this work Condition S. The
class of such Schr\"{o}dinger equations is really wide. We show that if the
potential in the Schr\"{o}dinger equation is spherically symmetric or it is a
function of one cartesian, or parabolic, or elliptic variable, the
corresponding Schr\"{o}dinger equation belongs to this class. Moreover, the
above mentioned cases are only some few examples. In general the
Schr\"{o}dinger equation belonging to the class must not necessarily admit
separation of variables.

The main result of the present work is a relatively simple procedure which
allows us to construct explicitly a locally complete system of solutions of
the Schr\"{o}dinger equation by one known particular solution. Here the local
completeness is understood in the sense that any solution of the
Schr\"{o}dinger equation can be approximated arbitrarily closely by a linear
combination of functions from this system in a neighborhood of any point of
the domain of interest. The global completeness is an open question
nevertheless the possibility to obtain a sequence of exact solutions with such
a special property as the local completeness can be useful in different
applications including qualitative analysis of solutions and numerical
solution of boundary value problems. The main result is based on a chain of
observations some of them representing independent interest.

First of all we observe (Subsection 3.1) that given a particular solution of
the stationary two-dimensional Schr\"{o}dinger equation the corresponding
Schr\"{o}dinger operator can be factorized just as in a one-dimensional
situation. In the considered two-dimensional case the factorizing terms are
operators $\partial_{\overline{z}}+\frac{\partial_{z}f_{0}}{f_{0}}C$ and
$\partial_{z}-\frac{\partial_{z}f_{0}}{f_{0}}C$, where $f_{0}$ is a particular
solution of the Schr\"{o}dinger equation
\begin{equation}
\left(  -\Delta+\nu\right)  f=0 \label{SchrIntr}%
\end{equation}
and $C$ is the complex conjugation operator. This observation gives us a
simple relation between the Schr\"{o}dinger equation and the Vekua equation%
\begin{equation}
\left(  \partial_{\overline{z}}+\frac{\partial_{z}f_{0}}{f_{0}}C\right)  w=0.
\label{VekIntr1}%
\end{equation}
Every solution of one of these equations can be transformed into a solution of
the other and vice versa.

Next we show that solutions of this equation are closely related to solutions
of another Vekua equation%
\begin{equation}
\left(  \partial_{\overline{z}}-\frac{\partial_{\overline{z}}f_{0}}{f_{0}%
}C\right)  W=0\label{VekIntr2}%
\end{equation}
which we call the main Vekua equation. For this equation we always have a
generating pair $(F,G)$ in explicit form and the $(F,G)$-derivative of $W$
(the operation introduced by L. Bers) is a solution of (\ref{VekIntr1}). The
$(F,G)$-antiderivative of $w$ is a solution of (\ref{VekIntr2}). Moreover, the
real part of $W$ is necessarily a solution of (\ref{SchrIntr}) and the
imaginary part of $W$ is a solution of another Schr\"{o}dinger equation with
the potential $\left(  -\nu+2\left(  \frac{\left\vert \nabla f_{0}\right\vert
}{f_{0}}\right)  ^{2}\right)  $ which is precisely the potential which would
be expected to obtain after the Darboux transformation (see, e.g., \cite{MS}).
We obtain a transformation which allows us to construct the imaginary part of
$W$ by its real part and vice versa obtaining in this way an analogue of the
Darboux transformation for the two-dimensional Schr\"{o}dinger equation. Here
we should say that this transformation is not yet a long-sought definitive
solution of the problem of a multidimensional generalization of the
one-dimensional Darboux transformation (see, e.g., \cite{Sabatier}) because it
is not clear how to include in our consideration the eigenvalues of the
operator. Nevertheless it is a certain progress in generalizing the Darboux
transformation and deserves more attention.

Let $f_{0}$ be a function of some variable $\rho:$ $f_{0}=f_{0}(\rho)$ such
that the expression $\Delta\rho/\left\vert \nabla\rho\right\vert ^{2}$ is a
function of $\rho$. We denote it by $s(\rho)=\frac{\Delta\rho}{\left\vert
\nabla\rho\right\vert ^{2}}$ and say that $f_{0}$ satisfies Condition S. We
show that under this condition for equation (\ref{VekIntr2}) the formal powers
in the sense of L. Bers can be constructed explicitly. It should be noted that
formal powers play a part analogous to that of powers of the independent
variable $z$ in classical analytic function theory. They give us analogues of
Taylor series expansions for solutions of (\ref{VekIntr2}), and locally any
solution of (\ref{VekIntr2}) can be approximated arbitrarily closely by a
finite number of first members of its Taylor series.

As was explained above, the real parts of formal powers are solutions of the
Schr\"{o}dinger equation (\ref{SchrIntr}) and therefore similarly to real
parts of powers of $z$ which are very important in theory of harmonic
functions and are widely used for numerical solution of boundary value
problems for the Laplace equation, the real parts of formal powers give us a
locally complete system of solutions.

For the sake of simplicity we consider the Schr\"{o}dinger equation with a
real valued potential, and in the last section we explain how our results can
be generalized to the case of a complex valued potential.

\bigskip

\section{Some definitions and results from pseudoanalytic function theory}

This section is based on the results presented in \cite{Berskniga} and
\cite{BersStat}. Let $\Omega$ be a domain in $\mathbf{R}^{2}$. Throughout the
whole paper we suppose that $\Omega$ is a simply connected domain.

\subsection{Generating pairs, derivative and antiderivative}

\begin{definition}
A pair of complex functions $F$ and $G$ possessing in $\Omega$ partial
derivatives with respect to the real variables $x$ and $y$ is said to be a
generating pair if it satisfies the inequality
\[
\operatorname{Im}(\overline{F}G)>0\qquad\text{in }\Omega.
\]

\end{definition}

Denote $\partial_{\overline{z}}=\frac{\partial}{\partial x}+i\frac{\partial
}{\partial y}$ and $\partial_{z}=\frac{\partial}{\partial x}-i\frac{\partial
}{\partial y}$ (usually these operators are introduced with the factor $1/2$,
nevertheless here it is somewhat more convenient to consider them without it).
The following expressions are known as characteristic coefficients of the pair
$(F,G)$
\[
a_{(F,G)}=-\frac{\overline{F}G_{\overline{z}}-F_{\overline{z}}\overline{G}%
}{F\overline{G}-\overline{F}G},\qquad b_{(F,G)}=\frac{FG_{\overline{z}%
}-F_{\overline{z}}G}{F\overline{G}-\overline{F}G},
\]

\[
A_{(F,G)}=-\frac{\overline{F}G_{z}-F_{z}\overline{G}}{F\overline{G}%
-\overline{F}G},\qquad B_{(F,G)}=\frac{FG_{z}-F_{z}G}{F\overline{G}%
-\overline{F}G},
\]
where the subindex $\overline{z}$ or $z$ means the application of
$\partial_{\overline{z}}$ or $\partial_{z}$ respectively.

Every complex function $W$ defined in a subdomain of $\Omega$ admits the
unique representation $W=\phi F+\psi G$ where the functions $\phi$ and $\psi$
are real valued. Sometimes it is convenient to associate with the function $W$
the function $\omega=\phi+i\psi$. The correspondence between $W$ and $\omega$
is one-to-one.

The $(F,G)$-derivative $\overset{\cdot}{W}=\frac{d_{(F,G)}W}{dz}$ of a
function $W$ exists and has the form
\begin{equation}
\overset{\cdot}{W}=\phi_{z}F+\psi_{z}G=W_{z}-A_{(F,G)}W-B_{(F,G)}\overline{W}
\label{FGder}%
\end{equation}
if and only if
\begin{equation}
\phi_{\overline{z}}F+\psi_{\overline{z}}G=0. \label{phiFpsiG}%
\end{equation}
This last equation can be rewritten in the following form%
\[
W_{\overline{z}}=a_{(F,G)}W+b_{(F,G)}\overline{W}%
\]
which we call the Vekua equation. Solutions of this equation are called
$(F,G)$-pseudoanalytic functions. If $W$ is $(F,G)$-pseudoanalytic, the
associated function $\omega$ is called $(F,G)$-pseudoanalytic of second kind.

\begin{remark}
The functions $F$ and $G$ are $(F,G)$-pseudoanalytic, and $\overset{\cdot}%
{F}\equiv\overset{\cdot}{G}\equiv0$.
\end{remark}

\begin{definition}
\label{DefSuccessor}Let $(F,G)$ and $(F_{1},G_{1})$ - be two generating pairs
in $\Omega$. $(F_{1},G_{1})$ is called \ successor of $(F,G)$ and $(F,G)$ is
called predecessor of $(F_{1},G_{1})$ if%
\[
a_{(F_{1},G_{1})}=a_{(F,G)}\qquad\text{and}\qquad b_{(F_{1},G_{1})}%
=-B_{(F,G)}\text{.}%
\]

\end{definition}

The importance of this definition becomes obvious from the following statement.

\begin{theorem}
\label{ThBersDer}Let $W$ be an $(F,G)$-pseudoanalytic function and let
$(F_{1},G_{1})$ be a successor of $(F,G)$. Then $\overset{\cdot}{W}$ is an
$(F_{1},G_{1})$-pseudoanalytic function.
\end{theorem}

\begin{definition}
\label{DefAdjoint}Let $(F,G)$ be a generating pair. Its adjoint generating
pair $(F,G)^{\ast}=(F^{\ast},G^{\ast})$ is defined by the formulas%
\[
F^{\ast}=-\frac{2\overline{F}}{F\overline{G}-\overline{F}G},\qquad G^{\ast
}=\frac{2\overline{G}}{F\overline{G}-\overline{F}G}.
\]

\end{definition}

The $(F,G)$-integral is defined as follows
\[
\int_{\Gamma}Wd_{(F,G)}z=\frac{1}{2}\left(  F(z_{1})\operatorname{Re}%
\int_{\Gamma}G^{\ast}Wdz+G(z_{1})\operatorname{Re}\int_{\Gamma}F^{\ast
}Wdz\right)
\]
where $\Gamma$ is a rectifiable curve leading from $z_{0}$ to $z_{1}$.

If $W=\phi F+\psi G$ is an $(F,G)$-pseudoanalytic function where $\phi$ and
$\psi$ are real valued functions then
\begin{equation}
\int_{z_{0}}^{z}\overset{\cdot}{W}d_{(F,G)}z=W(z)-\phi(z_{0})F(z)-\psi
(z_{0})G(z),\label{FGAnt}%
\end{equation}
and as $\overset{\cdot}{F}=\overset{}{\overset{\cdot}{G}=}0$, this integral is
path-independent and represents the $(F,G)$-antiderivative of $\overset{\cdot
}{W}$.

\subsection{Generating sequences and Taylor series in formal
powers\label{SubsectGenSeq}}

\begin{definition}
\label{DefSeq}A sequence of generating pairs $\left\{  (F_{m},G_{m})\right\}
$, $m=0,\pm1,\pm2,\ldots$ , is called a generating sequence if $(F_{m+1}%
,G_{m+1})$ is a successor of $(F_{m},G_{m})$. If $(F_{0},G_{0})=(F,G)$, we say
that $(F,G)$ is embedded in $\left\{  (F_{m},G_{m})\right\}  $.
\end{definition}

\begin{theorem}
Let \ $(F,G)$ be a generating pair in $\Omega$. Let $\Omega_{1}$ be a bounded
domain, $\overline{\Omega}_{1}\subset\Omega$. Then $(F,G)$ can be embedded in
a generating sequence in $\Omega_{1}$.
\end{theorem}

\begin{definition}
A generating sequence $\left\{  (F_{m},G_{m})\right\}  $ is said to have
period $\mu>0$ if $(F_{m+\mu},G_{m+\mu})$ is equivalent to $(F_{m},G_{m})$
that is their characteristic coefficients coincide.
\end{definition}

Let $W$ be an $(F,G)$-pseudoanalytic function. Using a generating sequence in
which $(F,G)$ is embedded we can define the higher derivatives of $W$ by the
recursion formula%
\[
W^{[0]}=W;\qquad W^{[m+1]}=\frac{d_{(F_{m},G_{m})}W^{[m]}}{dz},\quad
m=1,2,\ldots\text{.}%
\]

\begin{definition}
\label{DefFormalPower}The formal power $Z_{m}^{(0)}(a,z_{0};z)$ with center at
$z_{0}\in\Omega$, coefficient $a$ and exponent $0$ is defined as the linear
combination of the generators $F_{m}$, $G_{m}$ with real constant coefficients
$\lambda$, $\mu$ chosen so that $\lambda F_{m}(z_{0})+\mu G_{m}(z_{0})=a$. The
formal powers with exponents $n=1,2,\ldots$ are defined by the recursion
formula%
\begin{equation}
Z_{m}^{(n+1)}(a,z_{0};z)=(n+1)\int_{z_{0}}^{z}Z_{m+1}^{(n)}(a,z_{0}%
;\zeta)d_{(F_{m},G_{m})}\zeta.\label{recformula}%
\end{equation}

\end{definition}

This definition implies the following properties.

\begin{enumerate}
\item $Z_{m}^{(n)}(a,z_{0};z)$ is an $(F_{m},G_{m})$-pseudoanalytic function
of $z$.

\item If $a^{\prime}$ and $a^{\prime\prime}$ are real constants, then
$Z_{m}^{(n)}(a^{\prime}+ia^{\prime\prime},z_{0};z)=a^{\prime}Z_{m}%
^{(n)}(1,z_{0};z)+a^{\prime\prime}Z_{m}^{(n)}(i,z_{0};z).$

\item The formal powers satisfy the differential relations%
\[
\frac{d_{(F_{m},G_{m})}Z_{m}^{(n)}(a,z_{0};z)}{dz}=nZ_{m+1}^{(n-1)}%
(a,z_{0};z).
\]

\item The asymptotic formulas
\[
Z_{m}^{(n)}(a,z_{0};z)\sim a(z-z_{0})^{n},\quad z\rightarrow z_{0}%
\]
hold.
\end{enumerate}

Assume now that
\begin{equation}
W(z)=\sum_{n=0}^{\infty}Z^{(n)}(a,z_{0};z) \label{series}%
\end{equation}
where the absence of the subindex $m$ means that all the formal powers
correspond to the same generating pair $(F,G),$ and the series converges
uniformly in some neighborhood of $z_{0}$. It can be shown that the uniform
limit of pseudoanalytic functions is pseudoanalytic, and that a uniformly
convergent series of $(F,G)$-pseudoanalytic functions can be $(F,G)$%
-differentiated term by term. Hence the function $W$ in (\ref{series}) is
$(F,G)$-pseudoanalytic and its $r$th derivative admits the expansion
\[
W^{[r]}(z)=\sum_{n=r}^{\infty}n(n-1)\cdots(n-r+1)Z_{r}^{(n-r)}(a_{n}%
,z_{0};z).
\]
From this the Taylor formulas for the coefficients are obtained%
\begin{equation}
a_{n}=\frac{W^{[n]}(z_{0})}{n!}. \label{Taylorcoef}%
\end{equation}

\begin{definition}
Let $W(z)$ be a given $(F,G)$-pseudoanalytic function defined for small values
of $\left\vert z-z_{0}\right\vert $. The series%
\begin{equation}
\sum_{n=0}^{\infty}Z^{(n)}(a,z_{0};z) \label{Taylorseries}%
\end{equation}
with the coefficients given by (\ref{Taylorcoef}) is called the Taylor series
of $W$ at $z_{0}$, formed with formal powers.
\end{definition}

The Taylor series always represents the function asymptotically:%
\begin{equation}
W(z)-\sum_{n=0}^{N}Z^{(n)}(a,z_{0};z)=O\left(  \left\vert z-z_{0}\right\vert
^{N+1}\right)  ,\quad z\rightarrow z_{0}, \label{asympt}%
\end{equation}
for all $N$. This implies (since a pseudoanalytic function can not have a zero
of arbitrarily high order without vanishing identically) that the sequence of
derivatives $\left\{  W^{[n]}(z_{0})\right\}  $ determines the function $W$ uniquely.

If the series (\ref{Taylorseries}) converges uniformly in a neighborhood of
$z_{0}$, it converges to the function $W$.

\begin{theorem}
\label{ThConvPer} The formal Taylor expansion (\ref{Taylorseries}) of a
pseudoanalytic function in formal powers defined by a periodic generating
sequence converges in some neighborhood of the center.
\end{theorem}

We will use also the following observation from \cite[p. 140]{Bauer}.

\begin{proposition}
\label{PrBauer} \ Let $b$ be a complex function such that $b_{z}$ is real
valued, and let $W=u+iv$ be a solution of the equation
\[
W_{\overline{z}}=b\overline{W}.
\]
Then $u$ is a solution of the equation%
\begin{equation}
\partial_{\overline{z}}\partial_{z}u-(b\overline{b}+b_{z})u=0 \label{Bauer1}%
\end{equation}
and $v$ is a solution of the equation%
\begin{equation}
\partial_{\overline{z}}\partial_{z}v-(b\overline{b}-b_{z})v=0. \label{Bauer2}%
\end{equation}

\end{proposition}

\bigskip

\section{Relationship between generalized analytic functions and solutions of
the Schr\"{o}dinger equation}

\subsection{Factorization of the Schr\"{o}dinger operator}

It is well known that if $f_{0}$ is a nonvanishing particular solution of the
one-dimensional stationary Schr\"{o}dinger equation%
\[
\left(  -\frac{d^{2}}{dx^{2}}+\nu(x)\right)  f(x)=0
\]
then the Schr\"{o}dinger operator can be factorized as follows%
\[
\frac{d^{2}}{dx^{2}}-\nu(x)=\left(  \frac{d}{dx}+\frac{f_{0}^{\prime}}{f_{0}%
}\right)  \left(  \frac{d}{dx}-\frac{f_{0}^{\prime}}{f_{0}}\right)  .
\]
We start with a generalization of this result onto a two-dimensional
situation. Consider the equation%
\begin{equation}
\left(  -\Delta+\nu\right)  f=0 \label{Schrod}%
\end{equation}
in some domain $\Omega\subset\mathbf{R}^{2}$, where $\Delta=\frac{\partial
^{2}}{\partial x^{2}}+\frac{\partial^{2}}{\partial y^{2}}$, $\nu$ and $f$ are
real valued functions. We assume that $f$ is a twice continuously
differentiable function.

By $C$ we denote the complex conjugation operator.

\begin{theorem}
\label{Thfact}Let $f_{0}$ be a nonvanishing in $\Omega$ particular solution of
(\ref{Schrod}). Then for any real valued continuously twice differentiable
function $\varphi$ the following equality holds%
\begin{equation}
\left(  \Delta-\nu\right)  \varphi=\left(  \partial_{\overline{z}}%
+\frac{\partial_{z}f_{0}}{f_{0}}C\right)  \left(  \partial_{z}-\frac
{\partial_{z}f_{0}}{f_{0}}\right)  \varphi. \label{fact}%
\end{equation}

\end{theorem}

\begin{proof}
Consider%
\begin{align*}
\left(  \partial_{\overline{z}}+\frac{\partial_{z}f_{0}}{f_{0}}C\right)
\left(  \partial_{z}-\frac{\partial_{z}f_{0}}{f_{0}}\right)  \varphi &
=\Delta\varphi-\frac{\left\vert \partial_{z}f_{0}\right\vert ^{2}}{f_{0}^{2}%
}\varphi-\partial_{\overline{z}}\left(  \frac{\partial_{z}f_{0}}{f_{0}%
}\right)  \varphi\\
&  =\Delta\varphi-\frac{\Delta f_{0}}{f_{0}}\varphi=\left(  \Delta-\nu\right)
\varphi.
\end{align*}

\end{proof}

\begin{remark}
As $\varphi$ in (\ref{fact}) is a real valued function, we can add the
conjugation operator in the second first-order operator on the right-hand
side, and then (\ref{fact}) takes the form%
\[
\left(  \Delta-\nu\right)  \varphi=\left(  \partial_{\overline{z}}%
+\frac{\partial_{z}f_{0}}{f_{0}}C\right)  \left(  \partial_{z}-\frac
{\partial_{z}f_{0}}{f_{0}}C\right)  \varphi.
\]

\end{remark}

The operator $\partial_{z}-\frac{\partial_{z}f_{0}}{f_{0}}I$, where $I$ is the
identity operator, can be represented in the form%
\[
\partial_{z}-\frac{\partial_{z}f_{0}}{f_{0}}I=f_{0}\partial_{z}f_{0}^{-1}I.
\]
Let us introduce the following notation $P=f_{0}\partial_{z}f_{0}^{-1}I$. Due
to Theorem \ref{Thfact}, if $f_{0}$ is a nonvanishing solution of
(\ref{Schrod}), the operator $P$ transforms solutions of (\ref{Schrod}) into
solutions of the equation
\begin{equation}
\left(  \partial_{\overline{z}}+\frac{\partial_{z}f_{0}}{f_{0}}C\right)
w=0.\label{Vekua}%
\end{equation}

Note that the operator $\partial_{z}$ applied to a real valued function
$\varphi$ can be regarded as a kind of gradient, and if we know that
$\partial_{z}\varphi=\Phi$ in a whole complex plane or in a convex domain,
where $\Phi=\Phi_{1}+i\Phi_{2}$ is a given complex valued function such that
its real part $\Phi_{1}$ and imaginary part $\Phi_{2}$ satisfy the equation
\begin{equation}
\partial_{y}\Phi_{1}+\partial_{x}\Phi_{2}=0, \label{casirot}%
\end{equation}
then we can reconstruct $\varphi$ up to an arbitrary real constant $c$ in the
following way%
\begin{equation}
\varphi(x,y)=\int_{x_{0}}^{x}\Phi_{1}(\eta,y)d\eta-\int_{y_{0}}^{y}\Phi
_{2}(x_{0},\xi)d\xi+c \label{Antigr}%
\end{equation}
where $(x_{0},y_{0})$ is an arbitrary fixed point in the domain of interest.

By $A$ we denote the integral operator in (\ref{Antigr}):%
\[
A[\Phi](x,y)=\int_{x_{0}}^{x}\Phi_{1}(\eta,y)d\eta-\int_{y_{0}}^{y}\Phi
_{2}(x_{0},\xi)d\xi+c.
\]
Note that formula (\ref{Antigr}) can be easily extended to any simply
connected domain by considering the integral along an arbitrary rectifiable
curve $\Gamma$ leading from $(x_{0},y_{0})$ to $(x,y)$%
\[
\varphi(x,y)=\int_{\Gamma}\Phi_{1}dx-\Phi_{2}dy+c.
\]
Thus if $\Phi$ satisfies (\ref{casirot}), there exists a family of real valued
functions $\varphi$ such that $\partial_{z}\varphi=\Phi$, given by the formula
$\varphi=A[\Phi]$.

In a similar way we define the operator $\overline{A}$ corresponding to
$\partial_{\overline{z}}$:%
\[
\overline{A}[\Phi](x,y)=\int_{x_{0}}^{x}\Phi_{1}(\eta,y)d\eta+\int_{y_{0}}%
^{y}\Phi_{2}(x_{0},\xi)d\xi+c.
\]

Consider the operator $S=f_{0}Af_{0}^{-1}I$. It is clear that $PS=I$.

\begin{proposition}
\cite{KrPanalyt} \label{PrPS}Let $f_{0}$ be a nonvanishing particular solution
of (\ref{Schrod}) and $w$ be a solution of (\ref{Vekua}). Then the function
$f=Sw$ is a solution of (\ref{Schrod}).
\end{proposition}

\begin{proposition}
\cite{KrPanalyt} \label{PrSP}Let $f$ be a solution of (\ref{Schrod}). Then
\[
SPf=f+cf_{0}%
\]
where $c$ is an arbitrary real constant.
\end{proposition}

Theorem \ref{Thfact} together with Proposition \ref{PrPS} show us that
equation (\ref{Schrod}) is equivalent to the Vekua equation (\ref{Vekua}) in
the following sense. Every solution of one of these equations can be
transformed into a solution of the other equation and vice versa.

\subsection{The main Vekua equation}

Equation (\ref{Vekua}) is closely related to the following Vekua equation
\begin{equation}
\left(  \partial_{\overline{z}}-\frac{\partial_{\overline{z}}f_{0}}{f_{0}%
}C\right)  W=0. \label{Vekuaaux}%
\end{equation}
To see this let us observe that the pair of functions
\begin{equation}
F=f_{0}\quad\text{and\quad}G=\frac{i}{f_{0}} \label{genpair}%
\end{equation}
is a generating pair for (\ref{Vekuaaux}). Then the corresponding
characteristic coefficients $A_{(F,G)}$ and $B_{(F,G)}$ have the form%
\[
A_{(F,G)}=0,\quad\text{\quad}B_{(F,G)}=\frac{\partial_{z}f_{0}}{f_{0}},
\]
and the $(F,G)$-derivative according to (\ref{FGder}) is defined as follows%
\[
\overset{\cdot}{W}=W_{z}-\frac{\partial_{z}f_{0}}{f_{0}}\overline{W}=\left(
\partial_{z}-\frac{\partial_{z}f_{0}}{f_{0}}C\right)  W.
\]

Comparing $B_{(F,G)}$ with the coefficient in (\ref{Vekua}) and due to Theorem
\ref{ThBersDer} we obtain the following statement.

\begin{proposition}
\label{PrDer} Let $W$ be a solution of (\ref{Vekuaaux}). Then its
$(F,G)$-derivative, the function $w=\overset{\cdot}{W}$ is a solution of
(\ref{Vekua}).
\end{proposition}

This result can be verified also by a direct substitution.

According to (\ref{FGAnt}) and taking into account that
\[
F^{\ast}=-if_{0}\quad\text{and\quad}G^{\ast}=1/f_{0},
\]
the $(F,G)$-antiderivative has the form%
\begin{align}
\int_{z_{0}}^{z}w(\zeta)d_{(F,G)}\zeta &  =\frac{1}{2}\left(  f_{0}%
(z)\operatorname{Re}\int_{z_{0}}^{z}\frac{w(\zeta)}{f_{0}(\zeta)}d\zeta
-\frac{i}{f_{0}(z)}\operatorname{Re}\int_{z_{0}}^{z}if_{0}(\zeta
)w(\zeta)d\zeta\right) \nonumber\\
&  =\frac{1}{2}\left(  f_{0}(z)\operatorname{Re}\int_{z_{0}}^{z}\frac
{w(\zeta)}{f_{0}(\zeta)}d\zeta+\frac{i}{f_{0}(z)}\operatorname{Im}\int_{z_{0}%
}^{z}f_{0}(\zeta)w(\zeta)d\zeta\right)  , \label{antider}%
\end{align}
and we obtain the following statement.

\begin{proposition}
\label{PrAntider} Let $w$ be a solution of (\ref{Vekua}). Then the function
\[
W(z)=\int_{z_{0}}^{z}w(\zeta)d_{(F,G)}\zeta
\]
is a solution of (\ref{Vekuaaux}).
\end{proposition}

Let $\phi$ and $\psi$ be real valued functions. It is easy to see that the
function $W=\phi f_{0}+i\psi/f_{0}$ is a solution of (\ref{Vekuaaux}) if and
only if $\phi$ and $\psi$ satisfy the equation $\psi_{\overline{z}}-if_{0}%
^{2}\phi_{\overline{z}}=0$ which is equivalent to the system%
\[
\psi_{x}+f_{0}^{2}\phi_{y}=0,\qquad\psi_{y}-f_{0}^{2}\phi_{x}=0
\]
defining so called $p$-analytic functions (see \cite{Polozhy} and
\cite{KrPanalyt}) with $p=f_{0}^{2}$.

\begin{proposition}
\label{PrDarboux}Let $W$ be a solution of (\ref{Vekuaaux}). Then
$u=\operatorname{Re}W$ is a solution of (\ref{Schrod}) and
$v=\operatorname{Im}W$ is a solution of the equation%
\begin{equation}
\left(  \Delta+\nu-2\left(  \frac{\left\vert \nabla f_{0}\right\vert }{f_{0}%
}\right)  ^{2}\right)  v=0. \label{SchrodDarboux}%
\end{equation}

\end{proposition}

\begin{proof}
Observe that the coefficient $b=\frac{\partial_{\overline{z}}f_{0}}{f_{0}}$ in
(\ref{Vekuaaux}) satisfies the condition of Proposition \ref{PrBauer}:%
\[
b_{z}=\frac{\Delta f_{0}}{f_{0}}-\left(  \frac{\left\vert \partial
_{\overline{z}}f_{0}\right\vert }{f_{0}}\right)  ^{2}=\nu-\left(
\frac{\left\vert \partial_{\overline{z}}f_{0}\right\vert }{f_{0}}\right)
^{2}.
\]
Thus, according to Proposition \ref{PrBauer}, $u$ is a solution of
(\ref{Bauer1}) and $v$ is a solution of (\ref{Bauer2}). Calculating the
expressions $b\overline{b}+b_{z}=\nu$ and $b\overline{b}-b_{z}=2\left(
\frac{\left\vert \nabla f_{0}\right\vert }{f_{0}}\right)  ^{2}-\nu$ we finish
the proof.
\end{proof}

\begin{proposition}
\label{PrTransform}Let $u$ be a solution of (\ref{Schrod}). Then the function
\[
v\in\ker\left(  \Delta+\nu-2\left(  \frac{\left\vert \nabla f_{0}\right\vert
}{f_{0}}\right)  ^{2}\right)
\]
such that $W=u+iv$ is a solution of (\ref{Vekuaaux}), is constructed according
to the formula%
\begin{equation}
v=f_{0}^{-1}\overline{A}(if_{0}^{2}\partial_{\overline{z}}(f_{0}^{-1}u)).
\label{transfDarboux}%
\end{equation}
It is unique up to an additive term $cf_{0}^{-1}$ where $c$ is an arbitrary
real constant.

Given $v\in\ker\left(  \Delta+\nu-2\left(  \frac{\left\vert \nabla
f_{0}\right\vert }{f_{0}}\right)  ^{2}\right)  ,$ the corresponding $u\in
\ker\left(  \Delta-\nu\right)  $ can be constructed as follows%
\begin{equation}
u=-f_{0}\overline{A}(if_{0}^{-2}\partial_{\overline{z}}(f_{0}v))
\label{transfDarbouxinv}%
\end{equation}
up to an additive term $cf_{0}.$
\end{proposition}

\begin{proof}
Consider equation (\ref{Vekuaaux}). Let $W=\phi f_{0}+i\psi/f_{0}$ be its
solution. Then the equation
\begin{equation}
\psi_{\overline{z}}-if_{0}^{2}\phi_{\overline{z}}=0\label{seckind}%
\end{equation}
is valid. Note that if $u=\operatorname{Re}W$ then $\phi=u/f_{0}$. Given
$\phi$, $\psi$ is easily found from (\ref{seckind}):%
\[
\psi=\overline{A}(if_{0}^{2}\phi_{\overline{z}}).
\]
It can be verified that the expression $\overline{A}(if_{0}^{2}\phi
_{\overline{z}})$ makes sense, that is $\partial_{x}(f_{0}^{2}\phi
_{x})+\partial_{y}(f_{0}^{2}\phi_{y})=0.$

By Proposition \ref{PrDarboux} the function $v=f_{0}^{-1}\psi$ is a solution
of (\ref{SchrodDarboux}). Thus we obtain (\ref{transfDarboux}). Let us notice
that as the operator $\overline{A}$ reconstructs the scalar function up to an
arbitrary real constant, the function $v$ in the formula (\ref{transfDarboux})
is uniquely determined up to an additive term $cf_{0}^{-1}$ where $c$ is an
arbitrary real constant.

Equation (\ref{transfDarbouxinv}) is proved in a similar way.
\end{proof}

\begin{remark}
The potential in the Schr\"{o}dinger equation (\ref{SchrodDarboux}) has the
form of a potential obtained after the Darboux transformation (cf. \cite{MS},
\cite{ND}) and thus formulas (\ref{transfDarboux}) and (\ref{transfDarbouxinv}%
) can be considered as a two-dimensional analogue of the Darboux
transformation, though it is not clear how to include in our consideration the
eigenvalues of the operator.
\end{remark}

\begin{remark}
When $\nu\equiv0$ and $f_{0}\equiv1$, equalities (\ref{transfDarboux}) and
(\ref{transfDarbouxinv}) turn into the well known in complex analysis formulas
for constructing conjugate harmonic functions.
\end{remark}

\begin{remark}
Equation (\ref{Vekuaaux}) can be written as follows%
\begin{equation}
\left(  f_{0}\partial_{\overline{z}}f_{0}^{-1}P^{+}+if_{0}^{-1}\partial
_{\overline{z}}f_{0}P^{-}\right)  W=0\label{Vekuaproj}%
\end{equation}
where $P^{+}=\frac{1}{2}\left(  I+C\right)  $ and $P^{-}=\frac{1}{2i}\left(
I-C\right)  $.
\end{remark}

The form of the operator in (\ref{Vekuaproj}) suggests the following form of
an inverse operator%
\begin{equation}
H\Phi=\frac{1}{2}\left(  f_{0}\overline{A}\left(  f_{0}^{-1}\Phi\right)
+if_{0}^{-1}\overline{A}\left(  if_{0}\Phi\right)  \right)  \label{agregat}%
\end{equation}
where $\Phi$ must be such function that the expressions $\overline{A}\left(
f_{0}^{-1}\Phi\right)  $ and $\overline{A}\left(  if_{0}\Phi\right)  $ make sense.

\begin{proposition}
The function $W=H\Phi$ defined by (\ref{agregat}) is a solution of
(\ref{Vekuaproj}) and equivalently of (\ref{Vekuaaux}) if and only if
$\overline{\Phi}$ is a solution of (\ref{Vekua}).
\end{proposition}

\begin{proof}
Assume that $\Phi$ is such that the expressions $\overline{A}\left(
f_{0}^{-1}\Phi\right)  $ and $\overline{A}\left(  if_{0}\Phi\right)  $ make
sense, that is%
\begin{equation}
\partial_{y}\operatorname{Re}\left(  f_{0}^{-1}\Phi\right)  -\partial
_{x}\operatorname{Im}\left(  f_{0}^{-1}\Phi\right)  =0 \label{cond1}%
\end{equation}
and
\begin{equation}
\partial_{y}\operatorname{Re}\left(  if_{0}\Phi\right)  -\partial
_{x}\operatorname{Im}\left(  if_{0}\Phi\right)  =0. \label{cond2}%
\end{equation}
In this case let us substitute the function $W=H\Phi$ in (\ref{Vekuaproj}). We
have%
\begin{align*}
&  \frac{1}{2}\left(  f_{0}\partial_{\overline{z}}f_{0}^{-1}P^{+}+if_{0}%
^{-1}\partial_{\overline{z}}f_{0}P^{-}\right)  \left(  f_{0}\overline
{A}\left(  f_{0}^{-1}\Phi\right)  +if_{0}^{-1}\overline{A}\left(  if_{0}%
\Phi\right)  \right) \\
&  =\frac{1}{2}\left(  f_{0}\partial_{\overline{z}}\overline{A}\left(
f_{0}^{-1}\Phi\right)  +if_{0}^{-1}\partial_{\overline{z}}\overline{A}\left(
if_{0}\Phi\right)  \right)  =\frac{1}{2}(\Phi-\Phi)=0.
\end{align*}

Now let us prove that (\ref{cond1}) and (\ref{cond2}) are equivalent to the
fact that $\overline{\Phi}$ is a solution of (\ref{Vekua}). Denote
$\phi=\operatorname{Re}\Phi$ and $\psi=\operatorname{Im}\Phi$. Then
(\ref{cond1}) and (\ref{cond2}) can be written as follows%
\[
\partial_{x}\left(  \frac{\psi}{f_{0}}\right)  -\partial_{y}\left(  \frac
{\phi}{f_{0}}\right)  =0
\]
and
\[
\partial_{x}\left(  f_{0}\phi\right)  +\partial_{y}\left(  f_{0}\psi\right)
=0.
\]
The last two equalities are equivalent to the system%
\[
\partial_{x}\phi+\partial_{y}\psi=-\frac{\partial_{x}f_{0}}{f_{0}}\phi
-\frac{\partial_{y}f_{0}}{f_{0}}\psi,
\]%
\[
\partial_{x}\psi-\partial_{y}\phi=\frac{\partial_{x}f_{0}}{f_{0}}\psi
-\frac{\partial_{y}f_{0}}{f_{0}}\phi,
\]
which can be rewritten as the equation%
\[
\partial_{\overline{z}}\overline{\Phi}=-\frac{\partial_{z}f_{0}}{f_{0}}\Phi.
\]

\end{proof}

\begin{remark}
\label{RemAgregat} This proposition shows us that in the case of equation
(\ref{Vekuaaux}) the $(F,G)$-antiderivative can be calculated using
(\ref{agregat}). Indeed, let $W$ be a solution of (\ref{Vekuaaux}). Consider
its $(F,G)$-derivative $\overset{\cdot}{W}$ which is a solution of
(\ref{Vekua}) due to Proposition \ref{PrDer}. It can be written as follows%
\[
\overset{\cdot}{W}=f_{0}\partial_{z}\left(  f_{0}^{-1}u\right)  +if_{0}%
^{-1}\partial_{z}\left(  f_{0}v\right)  ,
\]
where $u=\operatorname{Re}W$ and $v=\operatorname{Im}W$. Consider%
\begin{align*}
HC\overset{\cdot}{W}  &  =\frac{1}{2}\left(  f_{0}\overline{A}\left(
f_{0}^{-1}C\overset{\cdot}{W}\right)  +if_{0}^{-1}\overline{A}\left(
if_{0}C\overset{\cdot}{W}\right)  \right) \\
&  =\frac{1}{2}f_{0}\overline{A}\left(  \partial_{\overline{z}}\left(
f_{0}^{-1}u\right)  \right)  -f_{0}\overline{A}\left(  if_{0}^{-2}%
\partial_{\overline{z}}\left(  f_{0}v\right)  \right) \\
&  +if_{0}^{-1}\overline{A}\left(  if_{0}^{2}\partial_{\overline{z}}\left(
f_{0}^{-1}u\right)  \right)  +if_{0}^{-1}\overline{A}\left(  \partial
_{\overline{z}}\left(  f_{0}v\right)  \right) \\
&  =\frac{1}{2}\left(  u+iv-f_{0}\overline{A}\left(  if_{0}^{-2}%
\partial_{\overline{z}}\left(  f_{0}v\right)  \right)  +if_{0}^{-1}%
\overline{A}\left(  if_{0}^{2}\partial_{\overline{z}}\left(  f_{0}%
^{-1}u\right)  \right)  \right)  .
\end{align*}
Frome here, due to Proposition \ref{PrDarboux} we obtain%
\[
HC\overset{\cdot}{W}=u+iv+c_{1}f_{0}+\frac{ic_{2}}{f_{0}},
\]
where $c_{1}$ and $c_{2}$ are arbitrary real constants.

Thus, application \ of the operator $HC$ to solutions of (\ref{Vekua}) gives
us exactly the same result as the $(F,G)$-antiderivative defined by
(\ref{antider}).
\end{remark}

\bigskip

\section{Taylor series in formal powers for pseudoanalytic functions and
solutions of the Schr\"{o}dinger equation}

In this section we show how for a quite ample class of potentials by one known
particular solution of (\ref{Schrod}) one can always construct an infinite
sequence of its solutions possessing the property of local completeness.

\subsection{Condition S}

\begin{lemma}
\label{LemEquivEqs} Let $\varphi$ be a nonvanishing analytic function. Then
solutions of the equations
\begin{equation}
W_{\overline{z}}=b\overline{W} \label{Vek1}%
\end{equation}
and
\begin{equation}
w_{\overline{z}}=\frac{\varphi}{\overline{\varphi}}b\overline{w} \label{Vek2}%
\end{equation}
are related in the following way. If $W$ is a solution of (\ref{Vek1}) then
$w=\varphi W$ is a solution of (\ref{Vek2}), and if $w$ is a solution of
(\ref{Vek2}) then $W=w/\varphi$ is a solution of (\ref{Vek1}).
\end{lemma}

The proof of this statement is obvious.

\begin{proposition}
\label{PrEquivEqs} Let $f_{0}$ be a function of some real variable $\rho$:
$f_{0}=f_{0}(\rho)$ such that for some real valued nonvanishing function
$\eta$ the equation
\begin{equation}
\partial_{\overline{z}}\left(  \eta\partial_{z}\rho\right)  =0\qquad\text{in
}\Omega\label{eqrho}%
\end{equation}
holds. Denote $\varphi=i\eta\rho_{z}$. Then
\[
\frac{\partial_{z}f_{0}}{f_{0}}=-\frac{\varphi}{\overline{\varphi}}%
\frac{\partial_{\overline{z}}f_{0}}{f_{0}},
\]
and if $W$ is a solution of (\ref{Vekuaaux}), the function $w=\varphi W$ is a
solution of (\ref{Vekua}) and vice versa, if $w$ is a solution of
(\ref{Vekua}), the function $W=w/\varphi$ is a solution of (\ref{Vekuaaux}).
\end{proposition}

\begin{proof}
Consider the expression $\frac{\partial_{z}f_{0}}{f_{0}}=\frac{f_{0}^{\prime
}\rho_{z}}{f_{0}}$. Observe that $\overline{\varphi}\rho_{z}=-\varphi
\rho_{\overline{z}}$. Then
\[
\frac{\partial_{z}f_{0}}{f_{0}}=-\frac{\varphi}{\overline{\varphi}}\frac
{f_{0}^{\prime}\rho_{\overline{z}}}{f_{0}}=-\frac{\varphi}{\overline{\varphi}%
}\frac{\partial_{\overline{z}}f_{0}}{f_{0}}.
\]
From (\ref{eqrho}) it is evident that $\varphi$ is analytic. Then denoting
$b=\frac{\partial_{\overline{z}}f_{0}}{f_{0}}$ we see that (\ref{Vekuaaux}) is
equation (\ref{Vek1}) from Lemma \ref{LemEquivEqs} and equation (\ref{Vekua})
is equation (\ref{Vek2}). Thus by Lemma \ref{LemEquivEqs} we obtain the result.
\end{proof}

Consider the following condition introduced in \cite{KrBers}.

\begin{condition}
\label{CondRho}(Condition S) Let $f_{0}$ be a function of some variable
$\rho:$ $f_{0}=f_{0}(\rho)$ such that $\frac{\Delta\rho}{\left\vert \nabla
\rho\right\vert ^{2}}$ is a function of $\rho$. We denote it by $s(\rho
)=\frac{\Delta\rho}{\left\vert \nabla\rho\right\vert ^{2}}$.
\end{condition}

The following proposition gives us a description of all possible solutions of
(\ref{eqrho}).

\begin{proposition}
\label{PrCritRho} For a real valued nontrivial function $\rho$ there exists a
real valued nonvanishing function $\eta$ such that (\ref{eqrho}) holds if and
only if $\rho$ satisfies Condition \ref{CondRho}.
\end{proposition}

\begin{proof}
Let $\rho$ satisfy Condition \ref{CondRho}. Denote $\eta=e^{-S}$ where $S$ is
the antiderivative of $s$ with respect to $\rho$. Consider%
\[
\partial_{\overline{z}}\left(  \eta\partial_{z}\rho\right)  =\partial
_{\overline{z}}\left(  e^{-S}\rho_{z}\right)  =-se^{-S}\left\vert \nabla
\rho\right\vert ^{2}+e^{-S}\Delta\rho=0.
\]

Now assume that for $\rho$ there exists a real valued nonvanishing function
$\eta$ such that (\ref{eqrho}) holds. Then $\Delta\rho+\frac{\eta
_{\overline{z}}}{\eta}\rho_{z}=0$ or in another form:
\[
\frac{\eta_{\overline{z}}}{\eta}=-\frac{\Delta\rho}{\left\vert \nabla
\rho\right\vert ^{2}}\rho_{\overline{z}}.
\]
If $\rho$ is harmonic then Condition \ref{CondRho} is obviously fulfilled, so
let us consider the opposite case assuming that $\rho$ is not harmonic. The
last equation can be written as follows%
\[
\nabla\ln\eta=\mu\nabla\rho,
\]
where $\mu=-\frac{\Delta\rho}{\left\vert \nabla\rho\right\vert ^{2}}$. In
order that the product $\mu\nabla\rho$ be a gradient it is necessary that
$\left[  \nabla\mu\times\nabla\rho\right]  =0$ which implies that $\mu
=\mu(\rho)$.
\end{proof}

Thus, due to Proposition \ref{PrCritRho} we have that the function $\varphi$
from Proposition \ref{PrEquivEqs} has the form $\varphi=ie^{-S}\rho_{z}$ where
$S(\rho)=\int\frac{\Delta\rho}{\left\vert \nabla\rho\right\vert ^{2}}d\rho$.

\subsection{Some examples of functions satisfying Condition S}

Examples of variables $\rho$ which satisfy Condition S are numerous and
important in applications. As was mentioned above any harmonic function $\rho$
fulfills the condition and obviously $\eta\equiv1$ in (\ref{eqrho}). That is,
for example, $\rho(x,y)=a_{1}x+a_{2}y,$ where $a_{1}$ and $a_{2}$ are
arbitrary real constants, or $\rho(x,y)=xy$ belong to that class.

An important example is $\rho(x,y)=r=\sqrt{x^{2}+y^{2}}$. In this case
$\frac{\Delta\rho}{\left\vert \nabla\rho\right\vert ^{2}}=\frac{1}{\rho}$.

The parabolic coordinate $\rho(x,y)=r+x$ also fulfills Condition S:
$\frac{\Delta\rho}{\left\vert \nabla\rho\right\vert ^{2}}=\frac{1}{2\rho}$.

Consider the elliptic coordinates $\mu$ and $\theta$:%
\[
x=\frac{a}{2}\cosh\mu\cos\theta\text{,\qquad}y=\frac{a}{2}\sinh\mu\sin\theta.
\]
It is convenient to consider the magnitudes%
\begin{align*}
r_{1} &  =\sqrt{\left(  x+\frac{a}{2}\right)  ^{2}+y^{2}}=\frac{a}{2}\left(
\cosh\mu+\cos\theta\right)  \text{,}\\
r_{2} &  =\sqrt{\left(  x-\frac{a}{2}\right)  ^{2}+y^{2}}=\frac{a}{2}\left(
\cosh\mu-\cos\theta\right)  .
\end{align*}
Let us verify Condition S for instance for the variable $\mu$. It is somewhat
easier to consider $\rho=a\cosh\mu$. If Condition S is fulfilled for $\rho$
then it is obviously true for $\mu$. We have $\rho=r_{1}+r_{2}$, and
\[
\rho_{z}=\frac{\overline{z}+a/2}{r_{1}}+\frac{\overline{z}-a/2}{r_{2}}.
\]
Then
\begin{align*}
\left\vert \nabla\rho\right\vert ^{2} &  =\left(  \frac{\overline{z}%
+a/2}{r_{1}}+\frac{\overline{z}-a/2}{r_{2}}\right)  \left(  \frac{z+a/2}%
{r_{1}}+\frac{z-a/2}{r_{2}}\right)  \\
&  =\frac{\rho^{2}-a^{2}}{r_{1}r_{2}}%
\end{align*}
and%
\[
\Delta\rho=\frac{\rho}{r_{1}r_{2}}.
\]
We obtain%
\[
\frac{\Delta\rho}{\left\vert \nabla\rho\right\vert ^{2}}=\frac{\rho}{\rho
^{2}-a^{2}}.
\]
Thus in all considered cases when $\rho$ is one of the cartesian coordinates,
when $\rho=r$, when $\rho$ is one of the parabolic coordinates or when $\rho$
is one of the elliptic coordinates Condition S is fulfilled. Moreover, as the
Laplacian admits separation of variables in all the mentioned coordinate
systems, then if the potential $\nu$ is a function of such $\rho$, there
exists a particular solution of (\ref{Schrod}) $f_{0}=f_{0}(\rho)$, and the
results of this section are applicable to all Schr\"{o}dinger equations with
potentials depending on such $\rho$.

We should emphasize first that these are only some examples which definitely
do not exhaust all interesting in applications situations that can be covered
by Condition S. And second, in order that such a solution $f_{0}=f_{0}(\rho)$
exist fulfilling Condition S, it is obviously not necessary that $\nu$ be a
function of $\rho$.

\subsection{Explicitely constructed generating sequence for the main Vekua
equation with $f_{0}$ satisfying Condition S}

In what follows we assume that $f_{0}$ is a nonvanishing solution of
(\ref{Schrod}) satisfying Condition S.

\begin{theorem}
\label{ThGenSeq} Let $\varphi=ie^{-S}\rho_{z}\neq0$ in $\Omega$. Then the
generating pair $(F,G)$ with $F=f_{0}$ and $G=i/f_{0}$ is embedded in the
generating sequence $(F_{m},G_{m})$, $m=0,\pm1,\pm2,\ldots$ with
$F_{m}=\varphi^{m}F$ and $G_{m}=\varphi^{m}G$.
\end{theorem}

\begin{proof}
First of all let us show that $(F_{m},G_{m})$ is a generating pair for
$m=\pm1,\pm2,\ldots$ . Indeed we have
\[
\operatorname{Im}(\overline{F}_{m}G_{m})=\operatorname{Im}(\overline{\varphi
}^{m}\overline{F}\varphi^{m}G)=\operatorname{Im}(\left\vert \varphi\right\vert
^{2m}\overline{F}G)>0.
\]

Taking into account that $\varphi_{\overline{z}}=0$ it is easy to obtain the
following equalities
\[
a_{(F_{m},G_{m})}=\left\vert \varphi\right\vert ^{2m}a_{(F,G)}\equiv0,
\]%
\[
b_{(F_{m},G_{m})}=\frac{\varphi^{m}}{\overline{\varphi}^{m}}b_{(F,G)}%
\]
and%
\[
B_{(F_{m-1},G_{m-1})}=\frac{\varphi^{m-1}}{\overline{\varphi}^{m-1}}%
B_{(F,G)}.
\]
We should verify the equality
\begin{equation}
b_{(F_{m},G_{m})}=-B_{(F_{m-1},G_{m-1})} \label{vsp}%
\end{equation}
which turns into the equality%
\[
\frac{\varphi}{\overline{\varphi}}b_{(F,G)}=-B_{(F,G)}.
\]
As $b_{(F,G)}=\frac{\partial_{\overline{z}}f_{0}}{f_{0}}$ and $B_{(F,G)}%
=\frac{\partial_{z}f_{0}}{f_{0}}$ by Proposition \ref{PrEquivEqs} we obtain
that (\ref{vsp}) is true. Thus the sequence $(F_{m},G_{m})$, $m=0,\pm
1,\pm2,\ldots$ satisfies the conditions of Definition \ref{DefSeq} and
therefore it is a generating sequence.
\end{proof}

This theorem opens the way for explicit construction of formal powers of any
order $n\geq0$ corresponding to the generating pair $(f_{0},i/f_{0})$ as well
as to any generating pair embedded in the sequence proposed in Theorem
\ref{ThGenSeq}. As a consequence L. Bers' theory of series expansion for
pseudoanalytic functions can be used in order to obtain explicitly Taylor
series in formal powers for solutions of the Schr\"{o}dinger equation
(\ref{Schrod}) due to Propositions \ref{PrDarboux} and \ref{PrTransform}.

Due to (\ref{asympt}) we have that any pseudoanalytic function $W$ can be
approximated at any point $z_{0}\in\Omega$ with an arbitrary precision by
first $N$ members of its Taylor series in formal powers. As any solution of
(\ref{Schrod}) is a real part of some pseudoanalytic function $W$ satisfying
(\ref{Vekuaaux}), it can also be approximated with arbitrary precision by the
sum of real parts of the first $N$ members of the Taylor series in formal
powers of the function $W$.

\begin{definition}
Let $u(z)$ be a given solution of (\ref{Schrod}) defined for small values of
$\left\vert z-z_{0}\right\vert $, and let $W(z)$ be a solution of
(\ref{Vekuaaux}) constructed according to Proposition \ref{PrTransform} such
that $\operatorname{Re}W=u$. The series
\begin{equation}
\sum_{n=0}^{\infty}\operatorname{Re}Z^{(n)}(a_{n},z_{0};z)
\label{TaylorSchrod}%
\end{equation}
with the coefficients given by (\ref{Taylorcoef}) is called the Taylor series
of $u$ at $z_{0}$, formed with formal powers.
\end{definition}

\begin{theorem}%
\begin{equation}
u(z)-\sum_{n=0}^{N}\operatorname{Re}Z^{(n)}(a_{n},z_{0};z)=O\left(  \left\vert
z-z_{0}\right\vert ^{N+1}\right)  ,\quad z\rightarrow z_{0}, \label{uappr}%
\end{equation}
for all $N$, and if the series (\ref{TaylorSchrod}) converges uniformly in a
neighborhood of $z_{0}$, it converges to the function $u$.
\end{theorem}

\begin{proof}
is a direct consequence of (\ref{asympt}).
\end{proof}

Due to Theorem \ref{ThGenSeq} we are able to construct $Z^{(n)}(a_{n}%
,z_{0};z)$ explicitly in many practically interesting cases. The simplest case
is when $f_{0}$ is a function of one cartesian variable: $f_{0}=f_{0}(y)$,
that is we consider the equation
\begin{equation}
-\Delta f(x,y)+\nu(y)f(x,y)=0\qquad\text{in }\Omega. \label{Schrody}%
\end{equation}

Let us make a useful observation. In this case $\rho_{z}=-i$ and $\varphi=1$.
Thus $(F_{m},G_{m})=(F,G)$, $m=0,\pm1,\pm2,\ldots$ , and obviously we have a
periodic generating sequence with the period $1$. Consequently, according to
Theorem \ref{ThConvPer} in the case under consideration we can guarantee not
only the approximation (\ref{asympt}) and (\ref{uappr}) but also the
convergence of the Taylor series in formal powers in some neighborhood of the
center. Let us consider the following example.

\begin{example}
Let $\Omega$ be the unit circle with center at the origin. Consider equation
(\ref{Schrody}) with $\nu(y)=6/\left(  y+1\right)  ^{2}$. One particular
solution depending on $y$ only can be chosen as follows $f_{0}(y)=\left(
y+1\right)  ^{3}$. It is easy to find another solution of (\ref{Schrody}). We
choose it in the following form $u(y)=\left(  y+1\right)  ^{-2}$. Next we
construct the function $v$ such that $W=u+iv$ be a solution of (\ref{Vekuaaux}%
). Using (\ref{transfDarboux}) we obtain $v=5x\left(  y+1\right)
^{-3}+c\left(  y+1\right)  ^{-3}$ where $c$ is an arbitrary real constant. We
choose it equal to zero, so $v=5x\left(  y+1\right)  ^{-3}$. It can be easily
verified that the function $W=\left(  y+1\right)  ^{-2}+5ix\left(  y+1\right)
^{-3}$ is indeed a solution of (\ref{Vekuaaux}) where $\frac{\partial
_{\overline{z}}f_{0}}{f_{0}}=3i(y+1)^{-1}$. Our aim is to find the Taylor
series in formal powers of the function $W$ at the origin.

We find that $W(0)=1$. Then using the definition of formal powers we find
$Z^{(0)}(1,0;z)=\left(  y+1\right)  ^{3}$. In order to construct $Z^{(1)}$ we
need to calculate $\overset{\cdot}{W}(0).$ We have $\overset{\cdot}%
{W}(z)=\partial_{z}W(z)+3i(y+1)^{-1}\overline{W}(z)=10i\left(  y+1\right)
^{-3}.$ Thus $\overset{\cdot}{W}(0)=10i$. In order to apply (\ref{recformula})
we find $Z^{(0)}(10i,0;z)=10i/\left(  y+1\right)  ^{3}$. Then using
(\ref{recformula}) and Remark \ref{RemAgregat} we have
\[
Z^{(1)}(10i,0;z)=HCZ^{(0)}(10i,0;z)=\left(  y+1\right)  ^{-2}-\left(
y+1\right)  ^{3}+5ix\left(  y+1\right)  ^{-3}.
\]

The function $\widetilde{W}(z)=Z^{(0)}(1,0;z)+Z^{(1)}(10i,0;z)$ should satisfy
the equality $W(z)-\widetilde{W}(z)=O(\left\vert z\right\vert ^{2})$ when
$z\rightarrow0$. We see that in fact $\widetilde{W}$ coincides with $W$ as
could be expected due to the fact that $\overset{\cdot}{W}$ is the generating
function $G$ multiplied by a real constant and hence $W^{[2]}\equiv0$.
\end{example}

Formal powers' Property 2 from Subsection \ref{SubsectGenSeq} together with
Theorem \ref{ThGenSeq} allows us to construct in explicit form a locally
complete system of solutions of (\ref{Vekuaaux}) and of (\ref{Schrod}) in the
following sense. Due to Theorem \ref{ThGenSeq} when $f_{0}$ fulfills Condition
S we \ are able to construct the formal powers $Z^{(n)}(1,z_{0};z)$ and
$Z^{(n)}(i,z_{0};z)$, $n=0,1,2,\ldots$ corresponding to any point $z_{0}%
\in\Omega$. Due to Property 2 of formal powers we have that $Z^{(n)}%
(a,z_{0};z)$ for any Taylor coefficient $a$ can be easily expressed through
$Z^{(n)}(1,z_{0};z)$ and $Z^{(n)}(i,z_{0};z)$. Thus for any solution $W$ of
(\ref{Vekuaaux}) there exists a linear combination of $Z^{(n)}(1,z_{0};z)$ and
$Z^{(n)}(i,z_{0};z)$, $n=0,1,2,\ldots N$ such that (\ref{asympt}) is valid.

Hence for any solution $u$ of (\ref{Schrod}) there exists a linear combination
of $\operatorname{Re}Z^{(n)}(1,z_{0};z)$ and $\operatorname{Re}Z^{(n)}%
(i,z_{0};z)$, $n=0,1,2,\ldots N$ such that (\ref{uappr}) holds.

Let us illustrate the procedure of construction of this locally complete
system on the following example.

\begin{example}
Let $\Omega$ be the unit circle with center at the origin and $\alpha$,
$\beta$ positive real constants greater than $1$. For the Schr\"{o}dinger
equation (\ref{Schrod}) with
\begin{equation}
\nu(x,y)=-\frac{1}{4}\left(  \frac{1}{(x+\alpha)^{2}}+\frac{1}{(y+\beta)^{2}%
}\right)  \label{PotEx2}%
\end{equation}
we have the particular solution $f_{0}(x,y)=\sqrt{(x+\alpha)(y+\beta)}$.
Denote $\rho=(x+\alpha)(y+\beta)$. Then Condition S is fulfilled and we have
that the function $\varphi$ from Theorem \ref{ThGenSeq} is defined as follows
$\varphi=ie^{-S}\rho_{z}=z+c$, where $c=\alpha+i\beta$. Let us construct the
first formal powers $Z^{(n)}(1,0;z)$ and $Z^{(n)}(i,0;z)$. By Definition
\ref{DefFormalPower} we have%
\[
Z^{(0)}(1,0;z)=\sqrt{\frac{(x+\alpha)(y+\beta)}{\alpha\beta}}\qquad
\text{and}\qquad Z^{(0)}(i,0;z)=i\sqrt{\frac{\alpha\beta}{(x+\alpha)(y+\beta
)}}.
\]
In order to construct $Z^{(1)}(1,0;z)$ and $Z^{(1)}(i,0;z)$ by formula
(\ref{recformula}) we need first $Z_{1}^{(0)}(1,0;z)$ and $Z_{1}^{(0)}%
(i,0;z)$. We have that $F_{1}=(z+c)\sqrt{(x+\alpha)(y+\beta)}$ and
$G_{1}=i(z+c)/\sqrt{(x+\alpha)(y+\beta)}$. Then $Z_{1}^{(0)}(1,0;z)=\lambda
_{1}F_{1}(z)+\mu_{1}G_{1}(z)$ with $\lambda_{1}$ and $\mu_{1}$ satisfying the
equality $\lambda_{1}F_{1}(0)+\mu_{1}G_{1}(0)=1$ and $Z_{1}^{(0)}%
(i,0;z)=\lambda_{2}F_{1}(z)+\mu_{2}G_{1}(z)$ with $\lambda_{2}$ and $\mu_{2}$
satisfying the equality $\lambda_{2}F_{1}(0)+\mu_{2}G_{1}(0)=i$. We obtain
that
\begin{equation}
\lambda_{1}=\frac{\sqrt{\alpha}}{\sqrt{\beta}(\alpha^{2}+\beta^{2}%
)}\text{,\qquad}\mu_{1}=-\frac{\sqrt{\alpha\beta^{3}}}{(\alpha^{2}+\beta^{2})}
\label{lambda1}%
\end{equation}
and
\begin{equation}
\lambda_{2}=\frac{\sqrt{\beta}}{\sqrt{\alpha}(\alpha^{2}+\beta^{2}%
)}\text{,\qquad}\mu_{2}=\frac{\sqrt{\alpha^{3}\beta}}{(\alpha^{2}+\beta^{2})}.
\label{lambda2}%
\end{equation}
Consider
\begin{align*}
Z^{(1)}(a,0;z)  &  =\int_{0}^{z}Z_{1}^{(0)}(a,0;\zeta)d_{(F,G)}\zeta\\
&  =\frac{1}{2}(\sqrt{(x+\alpha)(y+\beta)}\operatorname{Re}\int_{0}^{z}\left(
\lambda(\zeta+c)+\mu\frac{(\zeta+c)i}{(x^{\prime}+\alpha)(y^{\prime}+\beta
)}\right)  d\zeta\\
&  +\frac{i}{\sqrt{(x+\alpha)(y+\beta)}}\operatorname{Im}\int_{0}^{z}\left(
\lambda(\zeta+c)(x^{\prime}+\alpha)(y^{\prime}+\beta)+\mu(\zeta+c)i\right)
d\zeta)
\end{align*}
where $\lambda$ and $\mu$ are real numbers such that $\lambda F_{1}(0)+\mu
G_{1}(0)=a$ and $\zeta=x^{\prime}+iy^{\prime}$. We have
\[
\operatorname{Re}\int_{0}^{z}(\zeta+c)d\zeta=\int_{0}^{1}((x^{2}%
-y^{2})t+\alpha x-\beta y)dt=\frac{(x^{2}-y^{2})}{2}+\alpha x-\beta y,
\]%
\[
\operatorname{Re}\int_{0}^{z}\frac{(\zeta+c)id\zeta}{(x^{\prime}%
+\alpha)(y^{\prime}+\beta)}=-\int_{0}^{1}\frac{2xyt+\alpha y+\beta
x}{(xt+\alpha)(yt+\beta)}dt
\]%
\[
=-2xy\left(  \frac{\alpha}{x(\alpha y-\beta x)}\ln\left(  \frac{x+\alpha
}{\alpha}\right)  -\frac{\beta}{y(\alpha y-\beta x)}\ln\left(  \frac{y+\beta
}{\beta}\right)  \right)  -\frac{\alpha y+\beta x}{\alpha y-\beta x}\ln
\frac{\alpha\left(  y+\beta\right)  }{\beta\left(  x+\alpha\right)  }%
\]%
\[
=\ln\left(  \frac{\alpha\beta}{(x+\alpha)(y+\beta)}\right)  ,
\]
and
\[
\operatorname{Im}\int_{0}^{z}(\zeta+c)id\zeta=\operatorname{Re}\int_{0}%
^{z}(\zeta+c)d\zeta=\frac{(x^{2}-y^{2})}{2}+\alpha x-\beta y,
\]%
\[
\operatorname{Im}\int_{0}^{z}(\zeta+c)(x^{\prime}+\alpha)(y^{\prime}%
+\beta)d\zeta=\frac{\left(  xy\right)  ^{2}}{2}+(\alpha\beta+xy)(\alpha
y+\beta x)+\alpha\beta xy+\frac{(\alpha y+\beta x)^{2}}{2}.
\]
Then
\begin{align*}
Z^{(1)}(a,0;z)  &  =\frac{\sqrt{(x+\alpha)(y+\beta)}}{2}(\lambda(\frac
{x^{2}-y^{2}}{2}+\alpha x-\beta y)+\mu\ln\left(  \frac{\alpha\beta}%
{(x+\alpha)(y+\beta)}\right) \\
&  +\frac{i}{2\sqrt{(x+\alpha)(y+\beta)}}(\lambda(\frac{\left(  xy\right)
^{2}}{2}+(\alpha\beta+xy)(\alpha y+\beta x)+\alpha\beta xy+\frac{(\alpha
y+\beta x)^{2}}{2})\\
&  +\mu(\frac{x^{2}-y^{2}}{2}+\alpha x-\beta y)).
\end{align*}

One can check that for any $\lambda$ and $\mu$ the real part of this function
is indeed a solution of (\ref{Schrod}) with the potential (\ref{PotEx2}).
Substituting $\lambda$ and $\mu$ from (\ref{lambda1}) or (\ref{lambda2}) we
obtain $Z^{(1)}(1,0;z)$ and $Z^{(1)}(i,0;z)$ respectively.
\end{example}

\section{Complex potentials}

Our approach can also be applied to the Schr\"{o}dinger equation
(\ref{Schrod}) with $\nu$ being a complex function, though in this case
complex numbers become insufficient, and one should consider the bicomplex
generalization of the pseudoanalytic function theory.

Together with the imaginary unit $i$ let us consider another imaginary unit
$j$ such that $j^{2}=-1$ and $ij=ji$. Bicomplex numbers have the form
$z_{1}+z_{2}i$ where $z_{1}$ and $z_{2}$ can be considered as complex with
respect to the unit $j$: $z_{1,2}=x_{1,2}+jy_{1,2}$, $x_{1,2}$ and $y_{1,2}$
being real numbers. Bicomplex numbers obviously form a commutative algebra
which contains a subset of zero divisors.

Now we assume that $\nu=\nu_{1}+j\nu_{2}$ where $\nu_{1}$ and $\nu_{2}$ are
real valued functions. The factorization (\ref{fact}) remains valid for
$\varphi=\varphi_{1}+j\varphi_{2}$ as well as all the results of the present work.

\section{Conclusions}

In the present work we considered the real stationary two-dimensional
Schr\"{o}dinger equation. With the aid of any its particular solution we
construct the Vekua equation possessing the following special property. The
real parts of its solutions are solutions of the original Schr\"{o}dinger
equation and the imaginary parts are solutions of an associated
Schr\"{o}dinger equation with a potential having the form of a potential
obtained after the Darboux transformation. After having applied L. Bers'
approach to this Vekua equation we obtained a locally complete system of
solutions of the original Schr\"{o}dinger equation which can be constructed
explicitly for an ample class of Schr\"{o}dinger equations, namely when the
Schr\"{o}dinger equation admits a particular solution satisfying the proposed
Condition S. We established that in such special cases as of the potential
being a function of one cartesian, spherical, parabolic or elliptic variable
the condition is fulfilled. We gave some examples of application of the
proposed procedure for obtaining a locally complete system of solutions of the
Schr\"{o}dinger equation. The procedure is algorithmically simple and can be
implemented with the aid of a computer system of symbolic or numerical
calculation. The obtained system of solutions is a good candidate for
numerical analysis of boundary value problems for the Schr\"{o}dinger equation.

\bigskip\textbf{Acknowledgement}

The author wishes to express his gratitude to CONACYT for supporting this work
via the research project 43432.

\end{document}